\documentclass[12pt]{spieman}  % 12pt font required by SPIE;
\usepackage{amsmath,amsfonts,amssymb}
\usepackage{graphicx}
\usepackage{setspace}
\usepackage{tocloft}

\title{Astro-comb calibrator and spectrograph characterization using a turn-key laser frequency comb}

\author[a,*]{Aakash Ravi}
\author[b]{David F. Phillips}
\author[c]{Matthias Beck}
\author[d]{Leopoldo L. Martin}
\author[d]{Massimo Cecconi}
\author[d]{Adriano Ghedina}
\author[d]{Emilio Molinari}
\author[c]{Albrecht Bartels}
\author[b]{Dimitar Sasselov}
\author[b]{Andrew Szentgyorgyi}
\author[a,b]{Ronald L. Walsworth}

\affil[a]{Department of Physics, Harvard University, Cambridge, MA 02138, USA}
\affil[b]{Harvard-Smithsonian Center for Astrophysics, 60 Garden St., Cambridge, MA 02138, USA}
\affil[c]{Laser Quantum GmbH, Max-Stromeyer-Str. 116, 78467 Konstanz, Germany}
\affil[d]{INAF\textemdash Fundaci\'{o}n Galileo Galilei, 38712 Bre\~{n}a Baja, Spain}

\cftpagenumbersoff{figure}
\cftpagenumbersoff{table} 
\begin{document} 
\maketitle

\begin{abstract}
Using a turn-key Ti:sapphire femtosecond laser frequency comb, an off-the-shelf supercontinuum device, and Fabry-Perot mode filters, we report the generation of a 16 GHz frequency comb spanning a 90 nm band about a center wavelength of 566 nm. The light from this astro-comb is used to calibrate the HARPS-N astrophysical spectrograph for precision radial velocity measurements. The comb-calibrated spectrograph achieves a stability of $\sim$ 1 cm/s within half an hour of averaging time. We also use the astro-comb as a reference for measurements of solar spectra obtained with a compact telescope, and as a tool to study intrapixel sensitivity variations on the CCD of the spectrograph.   
\end{abstract}

% Include a list of up to six keywords after the abstract

% Include email contact information for corresponding author
{\noindent \footnotesize\textbf{*} \linkable{aravi@physics.harvard.edu} }

\begin{spacing}{1}   % use double spacing for rest of manuscript

\section{Introduction}

Finding and characterizing Earth-like planets orbiting Sun-like stars is one of the most challenging goals of modern radial velocity (RV) exoplanet science \cite{Fischer2016}. Such searches place very demanding requirements on the wavelength calibration of astrophysical spectrographs: specifically, detecting $\sim$ 10 cm/s RV shifts over the course of months to years, corresponding to sub-MHz changes in Doppler-broadened stellar absorption lines that are many GHz broad. Atomic emission lines from hollow cathode lamps and absorption lines from iodine vapor cells have been the workhorse calibration tools. However, these sources neither have uniform spectral coverage nor long-term stability.  A laser frequency comb (LFC) that is referenced to an atomic clock provides an excellent solution to this problem, as it provides a very large set of equispaced frequency markers with very accurately known absolute frequencies \cite{Li2008,Steinmetz2008,Braje2008}.

Although LFCs for the calibration of astrophysical spectrographs (``astro-combs'') have been successfully demonstrated \cite{Li2008,Braje2008,Glenday2015,Steinmetz2008,Wilken2010,Wilken2012,Quinlan2010,Ycas2012,Yi2016,McCracken2017}, they have not yet seen widespread adoption as primary calibrators in the astronomical community due to their complexity and cost. The operation of astro-combs has so far required significant laser expertise. It is therefore imperative to simplify the use of astro-combs \cite{Probst2014,McCracken2017} to make them viable for the next generation of high-precision RV measurements.

Astro-combs need to have a large mode spacing ($\gtrsim$ 10 GHz) to match the resolution of astrophysical spectrographs. Ti:sapphire-based astro-combs facilitate this because these lasers are available with intrinsically larger mode spacing compared to fiber lasers. But contrary to their fiber laser counterparts, these lasers require occasional realignment. Recently, however, this disadvantage has been overcome with the advent of turn-key Ti:sapphire LFCs.

Here, we demonstrate the simplification of the operation of an astro-comb by using a turn-key Ti:sapphire LFC, an off-the-shelf supercontinuum device, and existing Fabry-Perot mode filters. The laser is alignment-free, and therefore greatly simplifies the use of the astro-comb. We study the stability of our system, use it as a reference for measurements of solar spectra, and employ it to perform characterization of an astrophysical spectrograph.  We also discuss some potential future improvements to the system.

\section{Experimental setup}

\begin{figure}
\centering\includegraphics[width=\textwidth]{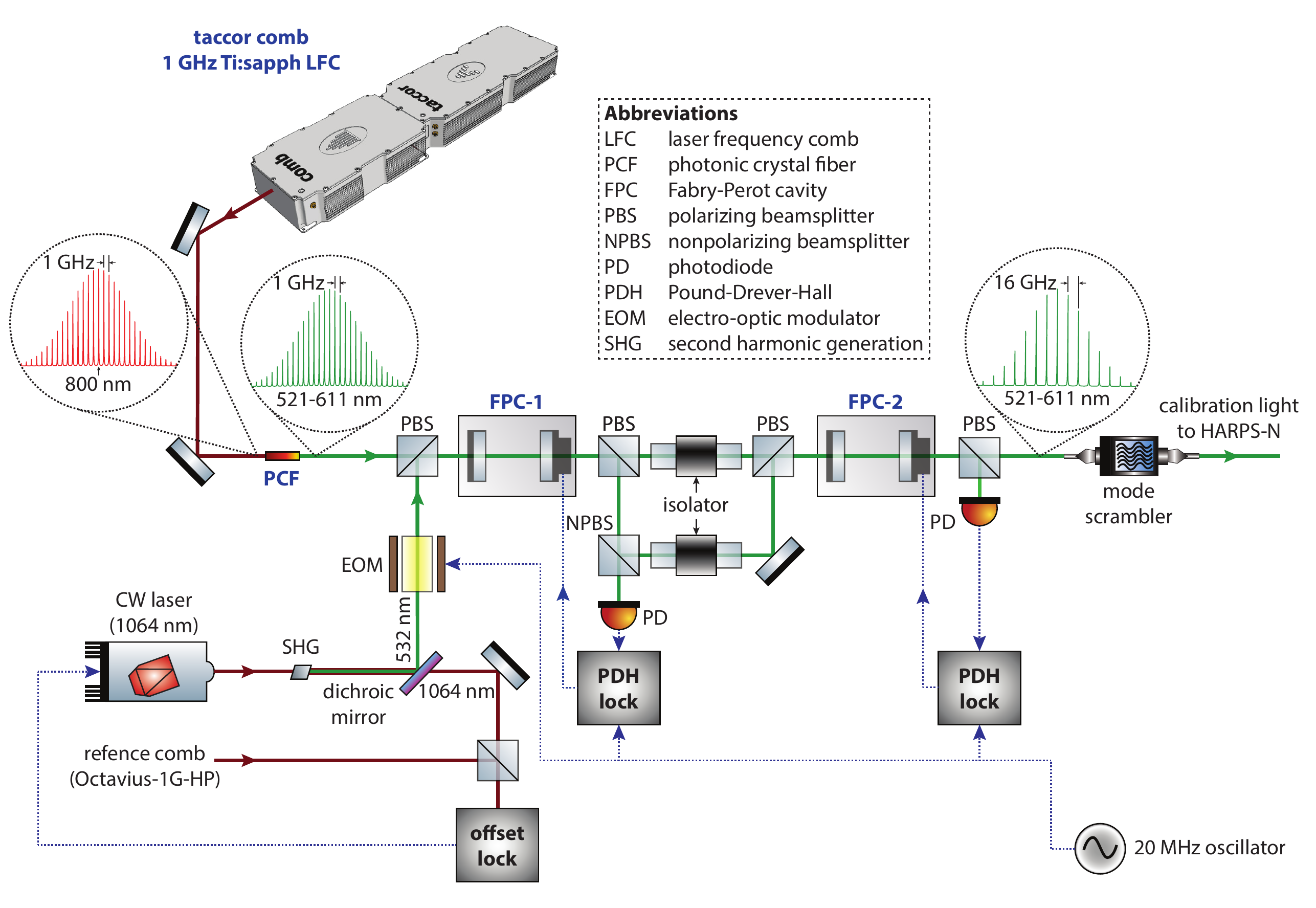}

\caption{\label{fig:schematic} Astro-comb block diagram showing a turn-key laser frequency comb that generates 1 GHz-spaced comb teeth about a center wavelength of 800 nm, a commercial photonic crystal fiber that coherently shifts the light into the visible wavelength range, and custom-built Fabry-Perot cavities that suppress 15 out of every 16 lines so as to match the resolution of the astrophysical spectrograph to be calibrated. (Abbreviations defined in figure; see text for more details.)}
\end{figure}

The astro-comb is located at the Telescopio Nazionale Galileo (TNG), on the island of La Palma in the Canary Islands, where it is used to calibrate the HARPS-N spectrograph \cite{Cosentino2012}. Figure~\ref{fig:schematic} shows a schematic of the experimental setup.

The astro-comb light source is the taccor comb (Laser Quantum), based on a turn-key 1 GHz Ti:sapphire mode-locked laser with a carrier-envelope offset (CEO) detection unit. The LFC operates at a center wavelength of about 800 nm and outputs $\sim$ 30 fs pulses at a repetition rate $f_{r}$ stabilized to a 1.000010870 GHz signal sourced from a RF synthesizer. The LFC carrier-envelope offset frequency $f_{0}$ is detected by sampling the optical output and sending it through a short length of nonlinear fiber to an \mbox{$f$-$2f$} detection unit, which directly locks the carrier-envelope offset frequency to 261.5 MHz. Both synthesizers are referenced to a GPS-disciplined 10 MHz Rb clock.

Approximately 300 mW of the source light is coupled into a supercontinuum device (NKT Photonics FemtoWHITE 800) to  spectrally broaden the 1 GHz repetition rate comb. This 12 cm-long aluminum-body device contains a photonic crystal fiber (PCF) with a 1.8 $\mu$m core diameter and a 750 nm zero dispersion wavelength. The device also features hole-collapsed, sealed fiber ends which are mounted in quartz ferrules. The output wavelengths relevant to calibrating the \mbox{HARPS-N} spectrograph are in the $\sim$500-600 nm range, but there is also a significant amount of light produced at longer wavelengths, up into the near infrared region. This light is then filtered by two 16 GHz free spectral range Fabry-Perot cavities in series, achieving \mbox{$>$ 40 dB} suppression of undesired comb teeth~\cite{Glenday2015}. The broadband cavities, which are based on zero group delay dispersion mirror pairs \cite{Chen2010}, are optimized for operation between 500-650 nm.  The cavity lengths are stabilized to a frequency-doubled 1064 nm CW single-frequency laser (JDSU NPRO 126N-1064-500) using a Pound-Drever-Hall scheme \cite{Drever1983} in transmission. As the cavities have residual dispersion, the frequency of the CW laser was empirically tuned to maximize the bandwidth transmitted by the cavities. This laser, in turn, is offset locked by heterodyning some of the 1064 nm light with a nearby tooth of an existing Ti:sapphire LFC (Menlo Systems Octavius-1G-HP) which is also referenced to the same GPS-disciplined signal sources as the taccor comb.  We would like to stress that, in the permanent setup, the reference laser will be offset locked directly to the taccor comb by heterodyning some of the frequency-doubled reference laser light at 532 nm with a nearby tooth of the PCF-broadened light from the taccor comb. Due to the limited time available for this demonstration however, we locked the reference laser  to the existing Octavius comb as in Ref. [\citen{Glenday2015}].

The spectrally broadened and filtered astro-comb light is coupled into a multimode fiber and sent through a mode scrambler~\cite{Glenday2015} to eliminate dynamic modal noise. The light is then sent via multimode fiber to HARPS-N \cite{Cosentino2012}, which is a high resolution \mbox{($R=$ 115,000)} cross-dispersed echelle spectrograph with spectral coverage from 380 nm to 690 nm. HARPS-N achieves $\sim$ few m/s RV stability prior to calibration by careful design, operation in vacuum, and temperature stabilization on the millikelvin level. Crucial to achieving sub-m/s RV observations is wavelength calibration, as well as simultaneous monitoring of potential calibration drifts while science exposures are performed. To this end, two input channels to HARPS-N are present: star light is injected into the ``science channel'' and calibration light into the ``reference channel.'' In the present work, we study the performance of the astro-comb by injecting its light into one or both channels of the HARPS-N spectrograph.

\section{Astro-comb characterization}

\begin{figure}
\centering\includegraphics[width=9.5cm]{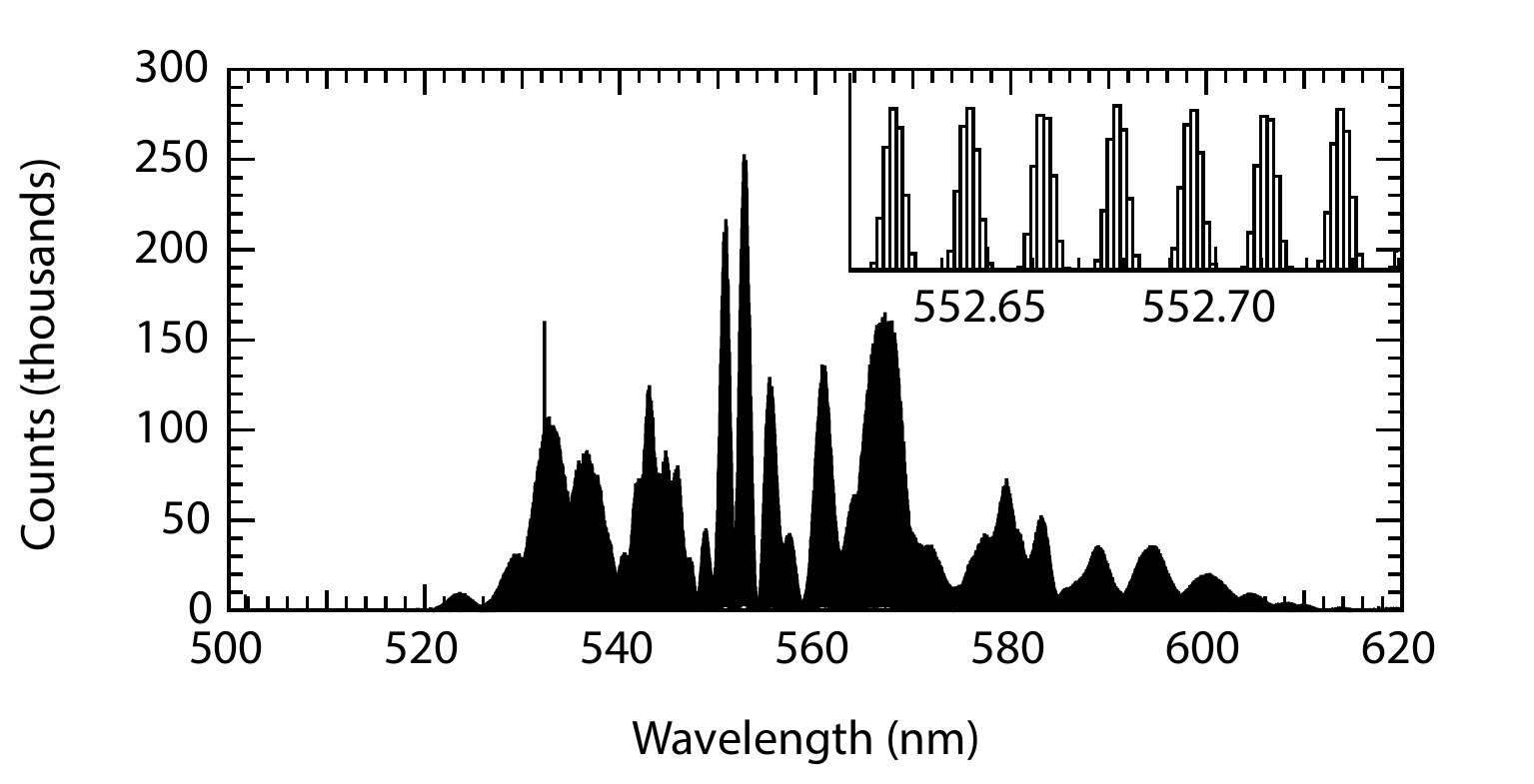}

\caption{\label{fig:spectrum}Astro-comb spectrum measured with HARPS-N spectrograph. The strong peak at 532 nm is the CW reference laser used to stabilize the filter cavity lengths. (Inset) At higher spectral resolution, individual comb teeth (spectral lines) are observed.}
\end{figure}

Figure~\ref{fig:spectrum} shows the broad spectrum of the astro-comb, as measured on the HARPS-N spectrograph. The -20 dB points of the spectrum lie at  521 and 611 nm. On a finer scale, individual comb spectral lines spaced by about 16 GHz are visible. 
The contrast of astro-comb peaks (peak height divided by background level) on HARPS-N is approximately 100. We believe this is due to long tails in the HARPS-N instrument profile as the contrast of the Octavius LFC-based astro-comb, observed with HARPS-N, is similar while its contrast on a high resolution Fourier transform spectrometer is \mbox{$>$10,000 ~\cite{Glenday2015}}. The strong modulation in the envelope of the spectrum is a result of the nonlinear processes in the supercontinuum device used for spectral
broadening. Improved spectral uniformity and even extending spectral coverage can be addressed by designing an optimized photonic crystal fiber for this LFC, and such efforts are currently underway. Alternatively, if the minimum number of counts in the present spectrum is satisfactory for calibration purposes, one could improve spectral uniformity by simply employing a lossy spectral flattening scheme with a spatial light modulator \cite{Probst2013}.

Ultimately, the bandwidth of our astro-comb is limited by the residual dispersion of filtering Fabry-Perot cavities \cite{Chang2012}. Extending cavity mirror spectral coverage is currently under investigation. An alternative approach that has been successfully demonstrated \cite{Probst2014,Quinlan2010,Ycas2012,Wilken2012} is to filter a narrow-band comb with a series of narrow-band high-finesse cavities and then perform the spectral broadening with a nonlinear fiber. This approach extends the bandwidth of the comb at the expense of system complexity, including added demands on Fabry-Perot cavity performance.

\begin{figure}
\centering\includegraphics[width=9.5cm]{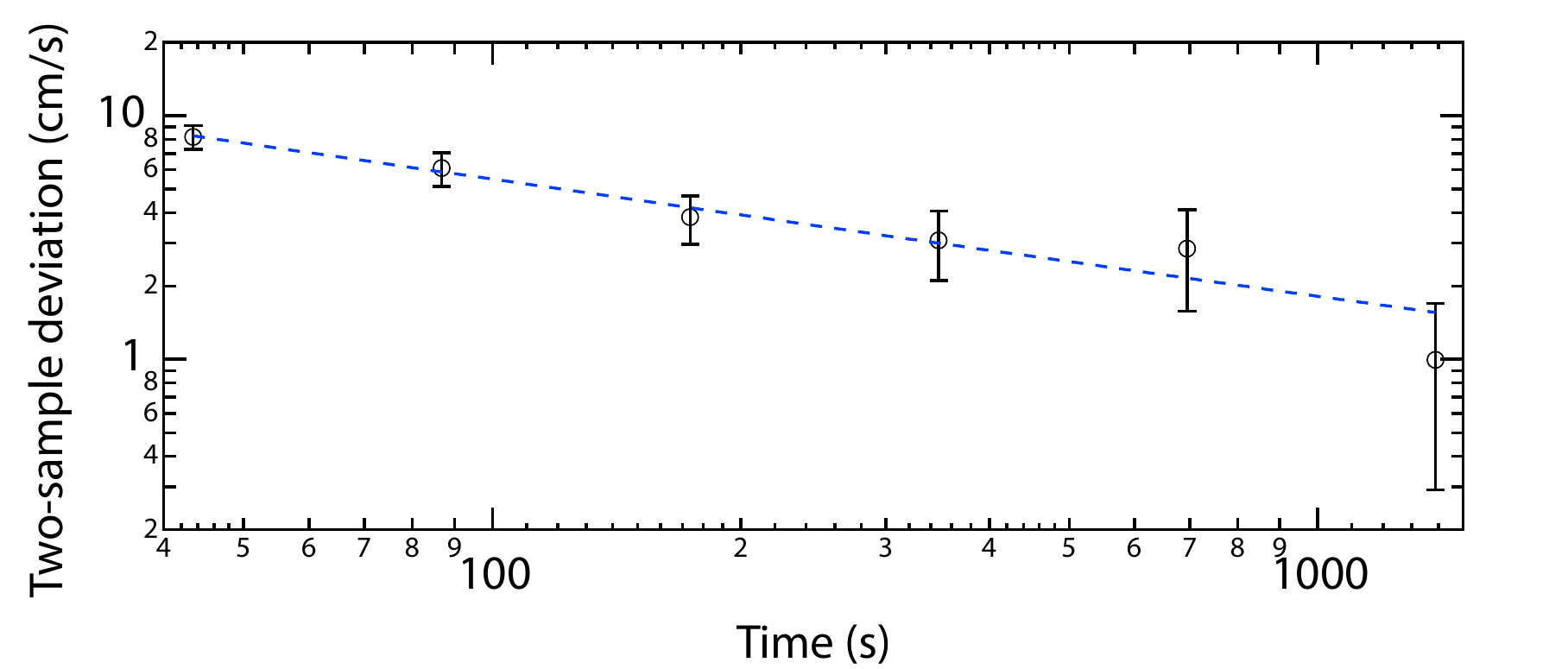}

\caption{\label{fig:allan}Two-sample deviation of the measured frequency stability
(in radial velocity units) of the astro-comb vs. averaging time. The black circles are the measured stability and dashed line is a fit consistent with the photon (shot) noise floor.  The overall duration of the measurements was approximately 1 hour.}
\end{figure}

We operated the astro-comb for several days, investigating its performance and stability.  To collect information about frequency stability, we injected astro-comb light into both channels of the HARPS-N spectrograph and monitored the deviation of their difference as a function of averaging time as shown in Fig.~\ref{fig:allan}. We consistently achieved RV sensitivity of nearly 1 cm/s at one half hour with no signs of spectrograph drift. Moreover, the two-sample deviation is consistent with the photon noise limit up to this point. This wavelength calibration is more than sufficient for RV detection of an Earth-analog exoplanet ($\sim$ 9 cm/s RV modulation).

\section{Comb-referenced solar spectra}

\begin{figure}
\centering\includegraphics[width=9cm]{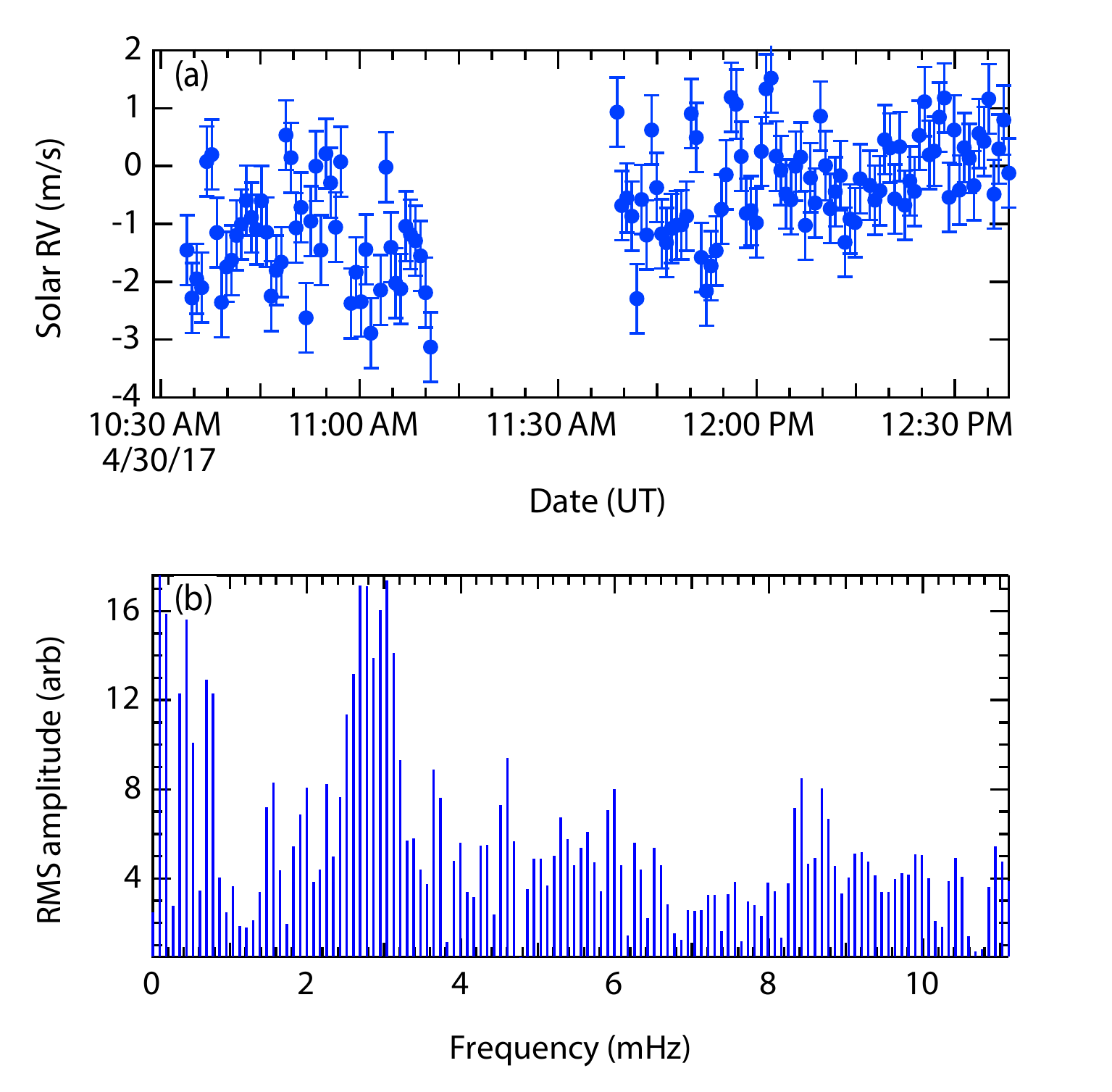}
%
%\centering\includegraphics[scale=0.152]{solarRV_PSD}
%
\caption{\label{fig:Solar_RV}(a) Comb-referenced solar RV
observations. (b) Power spectrum of the data.}
\end{figure}

As an example astronomical application, we performed solar spectral observations with \mbox{HARPS-N} referenced to the astro-comb. Using an automated compact solar telescope located at the TNG facility \cite{Dumusque2015}, we collect light from the full disk of the Sun and feed an integrating sphere to discard spatial information. This light is then injected into the science channel of the HARPS-N spectrograph, with the astro-comb light simultaneously injected into the reference channel. We took 20 second exposures for several hours, with a break due to clouds. In these short-term measurements, the astro-comb primarily provides a simultaneous reference as the radial velocity is only measured relative to the initial reference exposures for a few hours. We also derive a wavelength solution from the astro-comb spectrum injected in the science channel. Figure~\ref{fig:Solar_RV}a shows the difference between the observed RVs (calculated using a cross correlation technique with an empirical template and averaged across the orders of HARPS-N with significant astro-comb light) and the expected RVs from the JPL \emph{Horizons} ephemeris.  The power spectrum of these differences (Fig.~\ref{fig:Solar_RV}b) prominently shows the 5-minute (3 mHz) $p$-mode solar acoustic oscillations. However, there is also significant low-frequency noise present that is likely associated with granulation effects in the Sun~\cite{Dumusque2011}.

\section{Studies of intrapixel sensitivity variations of the spectrograph CCD}

A source of systematic error in RV measurements with an astrophysical spectrograph is non-uniformity in the detector. One example is intrapixel sensitivity variations \cite{Piterman2002,Toyozumi2005,Murphy2012} in the CCD. To study this effect, we took a sequence of exposures while shifting the astro-comb lines across half a pixel. Specifically, we shift the comb repetition rate $f_{r}$ since even a small change in $f_{r}$ gets magnified by the mode number $n$ through the relation defining the frequency of the $n^{\text{th}}$ line, $f_{n}=f_{0}+nf_{r}$. For example, a step of 3.2 kHz in the repetition rate amounts to moving a comb tooth across an entire pixel ($\sim$ 1.6 GHz). Note that we continue to take \mbox{$f_{r}=$ 1.000010870 GHz} to remain consistent with our discussion of the source comb; however, only every 16$^{\text{th}}$ line  appears on the spectrograph. 
%In our study, we take 10 upward steps of 160 Hz each, followed by 5 downward steps of 320 Hz each.

\begin{figure}
\centering\includegraphics[width=9.8cm]{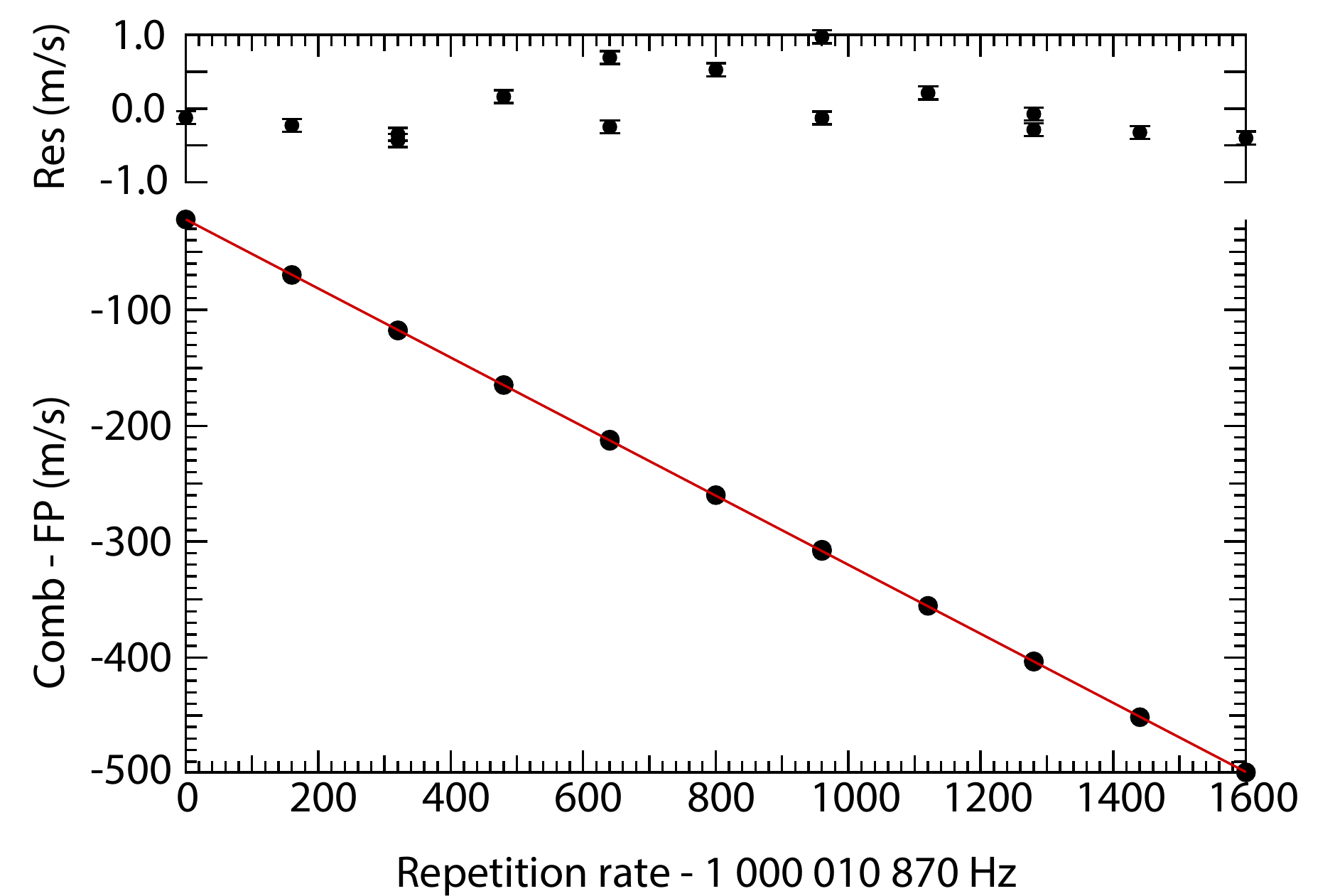}
\caption{\label{fig:frep-scan}Measured mean difference in astro-comb spectral lines relative to a white light FP spectrum (black circles) as a function of comb
repetition rate, with a linear fit (red line), and residuals  (top trace). The error bars of the residuals plot are the  statistical uncertainty of  \mbox{$\approx$ 10 cm/s} (see text).}
\end{figure}

\begin{figure}[t]
\centering\includegraphics[width=10.2cm]{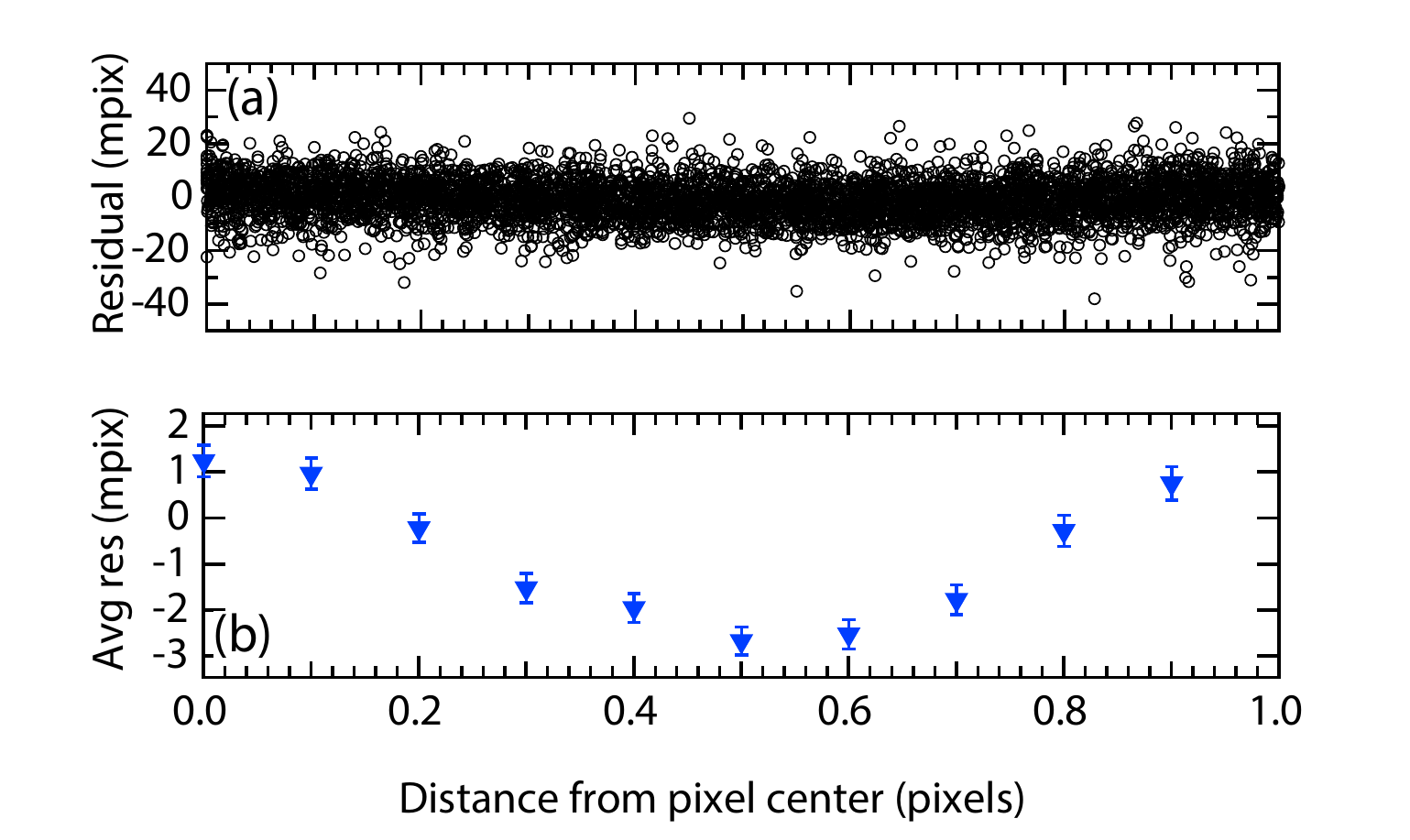}
\caption{\label{fig:peak-dev} (a) Difference between expected and actual astro-comb peak position, as a function of peak position inside a pixel. The uncertainty for each point is consistent with the $\sim$ 10 millipixel scatter observed. (b) Data shown in panel (a) averaged into bins with 0.1 pixel width.}
\end{figure}

To track the shifting astro-comb teeth as the repetition rate is changed, we inject comb light into the science channel of the spectrograph and light from a from a passively stabilized Fabry-Perot (FP) cavity illuminated by broadband white light (a laser-driven plasma source) into the reference channel as a fixed reference. Using a standard cross-correlation analysis, we calculate the mean shift in all the lines relative to the FP spectrum, assuming that the drift of the FP cavity is negligible during the entire measurement.  Figure~\ref{fig:frep-scan} shows the results of this analysis: a linear shift due to a change in repetition rate of the astro-comb of \mbox{0.298(2) m/s/Hz}, implying \mbox{$n\approx$ 523,000} which is a good estimate for the mean value of $n$ in $f_{n}=f_{0}+nf_{r}$. The upper panel of Fig.~\ref{fig:frep-scan} shows the residuals from the fit with error bars representing the statistical uncertainty of \mbox{$\approx$ 10 cm/s}.  We attribute  additional scatter in the residuals to systematic errors induced by changes in astro-comb power vs.\ wavelength as the repetition rate and reference laser are retuned. These variations do not affect the  results reported below and in Fig.~\ref{fig:peak-dev} as the analysis is henceforth done for each astro-comb peak independently.

%\begin{figure}
%\centering\includegraphics[width=11cm]{figure6}
%\caption{\label{fig:frac-part}Fractional CCD pixel values for 365 astro-comb spectral peaks in a single echelle order at $\sim$ 570 nm on the HARPS-N spectrograph. 16 consecutive exposures are shown. Color scale goes from black (0 = pixel center) to yellow (1 = neighboring pixel center). (Upper panel) In the first two-thirds of the exposures, the comb repetition rate ramps upward at 160 Hz per step; and in the last third  it ramps downward at 320 Hz per step. In so doing, comb spectral teeth are swept up and back down a distance of half a pixel.}
%\end{figure}

To search for intrapixel sensitivity variations, we fit all 365 astro-comb spectral peaks in a single echelle order at $\sim$ 570 nm (near the center of the astro-comb band) on the HARPS-N spectrograph to Gaussian profiles. As seen in the inset of Fig.~\ref{fig:spectrum}, typical peaks have a 3-4 pixel full width at half maximum. We model the mean intrapixel sensitivity variation of all the pixels in this order as follows: let $P_{i}^{j}$ be the pixel value of the fitted peak position of the $i^{\textrm{th}}$ peak in the $j^{\textrm{th}}$ exposure, where $i$ runs from 0 to 364, while $j$ runs from 0 to 15. $\mathrm{frac}\left(P_{i}^{j}\right)=P_{i}^{j}-\left \lfloor{P_{i}^{j}}\right \rfloor$ gives the fractional part of the peak position. Note that a value of 0 or 1 corresponds to pixel center while a value of 0.5 corresponds to a pixel boundary. We then compute a residual $R_{i}^{j}$ of the shift of the observed peak $i$ in exposure $j$ relative to the initial exposure compared to the expectation from shifting the repetition rate $f_{r}$ as
\begin{equation}
R_{i}^{j}=\left( P_{i}^{j}-P_{i}^{0} \right)- \Lambda\left(\Delta f_{r}, P_{i}^{j}\right) - \Delta_{\textrm{FP}}^{j,0} \left(P_{i}^{j}\right) 
\end{equation}
where $\Lambda\left(\Delta f_{r}, P_{i}^{j}\right)$ is the expected shift in pixels of the comb peak at $P_{i}^{j}$ due to the change in repetition rate $\Delta f_{r}$. The additional correction $\Delta_{\textrm{FP}}^{j,0} \left(P_{i}^{j}\right)$ is the local shift (i.e. interpolated to $P_{i}^{j}$) of the Fabry-Perot simultaneous reference between the initial exposure and the $j^{\textrm{th}}$ exposure. This parameter is extracted from a wavelength solution derived from the FP spectrum. Figure~\ref{fig:peak-dev}a shows $R_{i}^{j}$ as a function of $\mathrm{frac}\left(P_{i}^{j}\right)$, for all $i,j$. Averaging the data shown in Fig.~\ref{fig:peak-dev}a into bins of 0.1 pixels leaves a systematic residual at roughly the 5 millipixel level, as shown in Fig.~\ref{fig:peak-dev}b, which corresponds to $\sim$ 5 m/s in RV
units. Averaging over all the observed
lines in all the orders with different fractional CCD pixel values should reduce this systematic error below
the 1 m/s level. Key challenges for future work are to determine such intrapixel sensitivity variations across the full HARPS-N spectrum, and to mitigate its effects on astronomical RV observations.

\section{Conclusions}

In summary, we demonstrated an astro-comb
employing a turn-key mode-locked Ti:sapph laser, a commercially available
supercontinuum device, and existing Fabry-Perot mode filters. This astro-comb readily achieves RV stability of $\sim$ 1 cm/s within one half
hour averaging time. It is also successfully used as a reference for measurements of solar spectra and in the study of intrapixel sensitivity variations in
the CCD detector of the HARPS-N spectrograph. We are currently working to extend the spectral coverage, improve the intensity uniformity, and fully
automate the system (e.g. fiber alignment and filter cavity locks).

\acknowledgments
This research work was supported by the Harvard Origins of Life Initiative, the Smithsonian Astrophysical Observatory, NASA award number NNX16AD42G, NSF award number AST-1405606, and the Italian funding program ``Progetti Premiali'' WOW. A.R. was supported by a  postgraduate scholarship from the Natural Sciences and Engineering Research Council of Canada (NSERC).

%%%%% References %%%%%

\bibliography{references}   % bibliography data in references.bib
\bibliographystyle{spiejour}   % makes bibtex use spiejour.bst

%%%%% Biographies of authors %%%%%

\vspace{1ex}
\noindent Biographies and photographs of the authors are not available.

\listoffigures
%\listoftables

\end{spacing}
\end{document}